\documentclass[aps,prl,twocolumn,showpacs,superscriptaddress]{revtex4}
\usepackage{graphicx}
\begin{document}

\title{Paramagnetic spin excitations in insulating Rb$_{0.8}$Fe$_{1.6}$Se$_2$}
\author{Miaoyin Wang}
\affiliation{Department of Physics and Astronomy, The University of Tennessee, Knoxville,
Tennessee 37996-1200, USA }
\author{Xingye Lu}
\affiliation{Department of Physics and Astronomy, The University of Tennessee, Knoxville,
Tennessee 37996-1200, USA }
\affiliation{Beijing National Laboratory for Condensed Matter
Physics, Institute of Physics, Chinese Academy of Sciences, Beijing
100190, China}
\author{R. A. Ewings}
\affiliation{ISIS Facility, Rutherford Appleton Laboratory, Chilton, Didcot, Oxfordshire
OX11 0QX, UK}
\author{Leland W. Harriger}
\affiliation{Department of Physics and Astronomy, The University of Tennessee, Knoxville,
Tennessee 37996-1200, USA }
\author{Yu Song}
\affiliation{Department of Physics and Astronomy, The University of Tennessee, Knoxville,
Tennessee 37996-1200, USA }
\author{Scott V. Carr}
\affiliation{Department of Physics and Astronomy, The University of Tennessee, Knoxville,
Tennessee 37996-1200, USA }
\author{Chunhong Li}
\affiliation{Beijing National Laboratory for Condensed Matter
Physics, Institute of Physics, Chinese Academy of Sciences, Beijing
100190, China}
\author{Rui Zhang}
\affiliation{Beijing National Laboratory for Condensed Matter
Physics, Institute of Physics, Chinese Academy of Sciences, Beijing
100190, China}
\author{Pengcheng Dai}
\email{pdai@utk.edu}
\affiliation{Department of Physics and Astronomy, The University of Tennessee, Knoxville,
Tennessee 37996-1200, USA }
\affiliation{Beijing National Laboratory for Condensed Matter
Physics, Institute of Physics, Chinese Academy of Sciences, Beijing
100190, China}

\begin{abstract}
We use neutron scattering to study temperature dependent spin excitations in
insulating antiferromagnetic (AF) Rb$_{0.8}$Fe$_{1.6}$Se$_2$.  In the low-temperature AF state,  spin waves can be accurately
described by a local moment Heisenberg Hamiltonian.  On warming to around the
N$\rm \acute{e}$el temperature of $T_N=500$ K, low-energy ($E<30$ meV) 
paramagnetic spin excitations form Lorentzian-like quasielastic peaks 
centered at the AF wave vectors associated with spin waves, while high-energy ($E>50$ meV) spin excitations become heavily damped.  
Upon further warming to above the structural distortion temperature of $T_s=524$ K, the entire paramagnetic excitations become overdamped.  These results
suggest that AF Rb$_{0.8}$Fe$_{1.6}$Se$_2$ is not  a 
copper-oxide-like Mott insulator, and has less electron correlations compared with metallic 
 iron pnictides and iron chalcogenides.
\end{abstract}

\pacs{75.30.Ds, 75.50.Ee, 78.70.Nx, 29.30.Hs}

\maketitle
Since the discovery of antiferromagnetic (AF) order in the parent compounds of
iron pnictide superconductors \cite{Kamihara,cruz}, its microscopic origin and connection with
superconductivity has been an issue of controversy \cite{dai}.  One class of models, rooted in the semi-metallic nature of
these materials \cite{Kamihara}, argues that the collinear AF order in
the parent compounds such as BaFe$_2$As$_2$ \cite{qhuang} and SrFe$_2$As$_2$ \cite{jzhao08} is the spin-density-wave type originating from the nesting of itinerant electrons
between the hole and electron Fermi surfaces at $\Gamma$ and $M$ points in the Brillouin zone, respectively \cite{Hirschfeld}.  On the other hand, there are reasons to believe that
iron pnictides are not far away from a Mott insulator, where electron correlations
are important in determining the transport and magnetic perperties of these materials \cite{qmsi09}.
The discovery of insulating $A_y$Fe$_{1.6+x}$Se$_2$ ($A=$ K, Rb, Cs, Tl) near alkaline iron selenide
superconductors \cite{jguo,mhfang} provided a new opportunity to test whether the system is indeed a
Mott insulator similar to the insulating copper oxides \cite{ryu}, an AF semiconductor \cite{xwyan}, or
an insulator with 
coexisting itinerant and localized electronic state 
controlled by the Hund's rule coupling
\cite{yzyou,wgyin}.  Although the insulating $A_y$Fe$_{1.6+x}$Se$_2$ are isostructural with the metallic iron pnictides \cite{dai}, they form a $\sqrt{5}\times\sqrt{5}$
block AF structure with a large ($\sim$3.3 $\mu_B$ per Fe)
$c$-axis aligned moment and iron vacancy order (Fig. 1a), completely different from the collinear AF
structure of iron pnictides \cite{weibao,fye,mwang11}.

Using time-of-flight neutron spectroscopy, we showed previously that spin waves in insulating Rb$_{0.89}$Fe$_{1.58}$Se$_2$ can be accurately described by a local moment Heisenberg Hamiltonian \cite{mywang11}. For comparison, we note that there are still debates concerning whether a local moment Heisenberg Hamiltonian can appropriately model spin waves in
iron pnictides \cite{jzhao09,Diallo,ewings11,harriger11,sodiallo10,hpark,wysocki,ryu12}.  Moreover, recent spin wave measurements on 
iron chalcogenide Fe$_{1.1}$Te, which has a bicolinear AF structure and N$\rm \acute{e}$el temperature of $T_N=67$ K
\cite{Fang,Bao,Li,lipscombe}, suggest that the effective spin per Fe changes from $S\approx 1$ in the AF state to $S\approx 3/2$ in the paramagnetic state, 
much different from the expectation of
a conventional Heisenberg antiferromagnet \cite{igor}.  On the other hand, temperature dependent
paramagnetic scattering measurements
in metallic AF BaFe$_2$As$_2$ reveal that high-energy ($E>100$ meV) spin waves and 
the effective spin per Fe are
essentially unchanged for temperatures up to $2.1T_N$ \cite{harriger12}.  Given such diverse results in the parent compounds of iron-based superconductors, it is important to study the evolution of spin waves in a well-defined local moment Heisenberg system expected to be
close to a Mott transition \cite{ryu}.

\begin{figure}
\begin{center}
\includegraphics[width=0.88\linewidth]{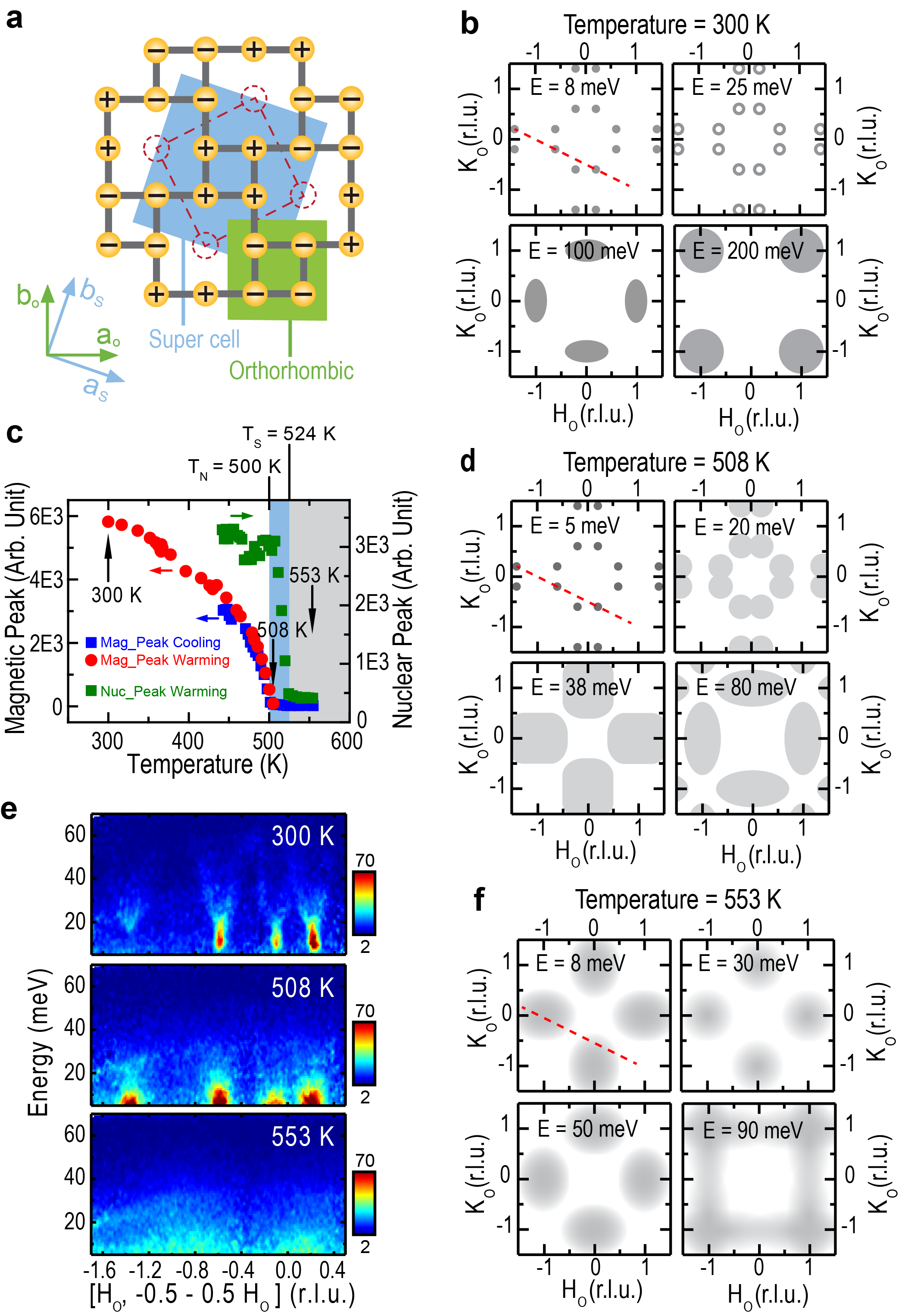}
\end{center}\caption{
(Color online)
(a) Nuclear and magnetic structures of irons in insulating Rb$_{0.8}$Fe$_{1.6}$Se$_2$.
The red dashed line square is the nuclear unit cell.  The blue shaded square is the magnetic unit cell.
The green shaded square is the orthorhombic
magnetic unit cell of iron pnictide  such as BaFe$_2$As$_2$ \cite{harriger11}.
(c) Temperature dependence of the magnetic and nuclear lattice distortion peaks obtained using a white incident neutron beam.
 The magnetic peak is at the in-plane wave vector $Q=(0.6, 0.2)$ rlu and
 the nuclear peak is at the $Q=(2, 0)$ rlu.  The $c$-axis momentum transfer is not well defined, and the data were obtained by integrating $L$ over a small
 region near the odd and even values, respectively \cite{mwang11}.
 The measurement shows that the N$\rm \acute{e}$el temperature is $T_N=500$ K and the structure transition temperature is about $T_s=524$ K.
(e) Spin wave energy versus wave vector projected along 
 the direction of the red-dashed lines in (b)(d)(f) or the $[H_o, -0.5-0.5H_o]$ direction)
 at $T=300$, 508, and 553 K, respectively.  The well-defined acoustic spin wave plumes are heavily damped at
508 K just above $T_N$, and essentially disappear at 553 K just above $T_s$. The vertical color bars are 
scattering intensity in mbarns sr$^{-1}$ meV$^{-1}$ f.u.$^{-1}$ (where f.u. is formula unit)
obtained by normalizing the magnetic scattering to a vanadium standard (with 20\% error) throughout the paper.
Compared with earlier spin wave work on ARCS at Spallation Neutron Source, Oak Ridge National Laboratory \cite{mywang11},
which has an error of 50\%, the present measurements on MAPS have more accurate absolute intensity normalization due to better detector calibration.
(b,d,f) Schematics of paramagnetic spin excitations at 300 K, 508 K, 553 K, respectively.
}
\label{Fig:fig1}
\end{figure}

In this paper, we report inelastic neutron scattering studies of paramagnetic spin excitations in
AF Rb$_{0.8}$Fe$_{1.6}$Se$_2$. In the low-temperature insulating state, Rb$_{0.8}$Fe$_{1.6}$Se$_2$ form a
$\sqrt{5}\times\sqrt{5}$ block AF structure with a large iron ordered moment
and iron vacancy order (Fig. 1a) \cite{weibao,fye,mwang11}.
Spin waves have three branches: one low-energy ($E\leq 80$ meV) acoustic spin wave branch stemming from the
block AF ordering wave vectors,
and two optical branches (at $E\approx 100$ and 200 meV, respectively) centered at wave vectors associated
with spin waves in iron pnictides (Fig. 1b) \cite{harriger11};
 and can be well described by a local moment
Heisenberg Hamiltonian \cite{mywang11}.
On warming to 508 K above $T_N=500$ K, the static AF order disappears but the lattice distortion induced
by the iron vacancy order persists (Fig. 1c).  Here, paramagnetic spin excitations at low-energies ($E\leq 30$ meV)
form Lorentzian-like quasielastic peaks centered at the block AF wave vectors, whereas
paramagnetic spin excitations at energies near optical spin waves are damped out (Fig. 1d).
Upon further warming to $T=1.05T_s=1.11T_N=553$ K, the $\sqrt{5}\times\sqrt{5}$
iron vacancy induced lattice distortion vanishes and the system becomes tetragonal with disordered iron vacancies \cite{mwang11}.
The low-energy ($<30$ meV) paramagnetic spin excitations are only weakly correlated at the AF ordering wave vectors
for iron pnictides.  Therefore, temperature dependence of spin waves in insulating Rb$_{0.89}$Fe$_{1.58}$Se$_2$
behaves like a local moment Heisenberg antiferromagnet, much different from that of metallic Fe$_{1.1}$Te \cite{igor}
and BaFe$_2$As$_2$ \cite{harriger12}.  These results indicate that insulating Rb$_{0.8}$Fe$_{1.6}$Se$_2$ has less electron correlations
and is not a copper-oxide-like Mott insulator.

\begin{figure}
\begin{center}
\includegraphics[width=0.8\linewidth]{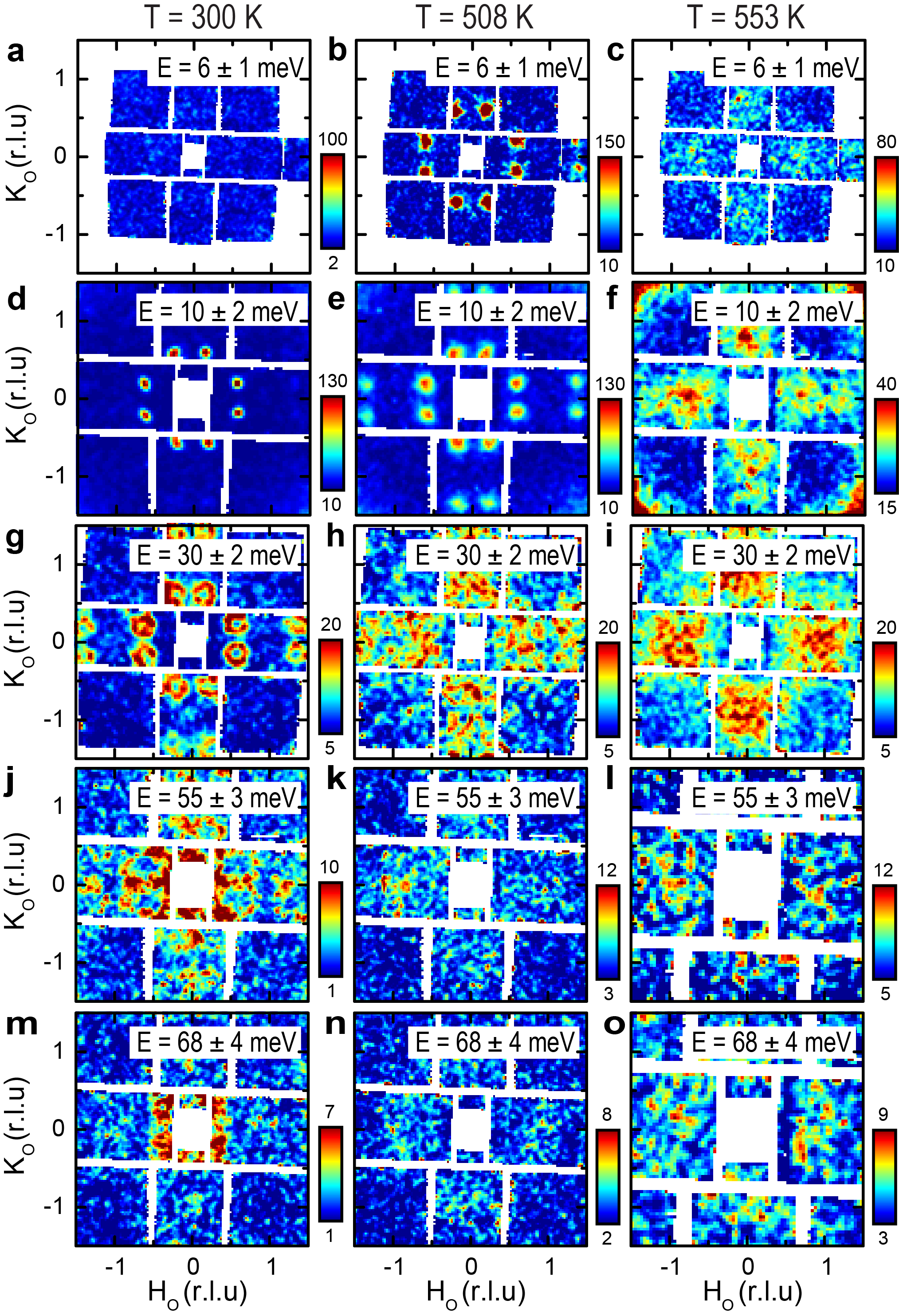}
\end{center}\caption{
(Color online) Wave vector and temperature dependence of acoustic spin wave and paramagnetic spin excitations at different energies for
Rb$_{0.8}$Fe$_{1.6}$Se$_2$. Spin wave and paramagnetic spin excitations in the $[H_o,K_o]$ scattering plane at energies
(a,b,c) $E=6\pm 1$, obtained with $E_i=35$ meV, corresponding to spin waves with $L=1.01$ in (a),  
(d,e,f) $E=10\pm 2$, (g,h,i) $E=30\pm 2$ meV, taken with $E_i=80$ meV,
(j,k,l) $E=55\pm3$, (m,n,o) $E=68\pm 4$ meV. Data in (j,k,m,n) are obtained with $E_i=140$ meV, 
while data in (l,o) are taken with $E_i=250$ meV.  In all cases, the incident beam is along the 
$c$-axis direction.
The left column is data at 300 K, the middle column is for 508 K, and the right column is at 553 K.  Energy resolution is about
10\% of the incident beam energy and decreases with increasing energy transfer.
}
\label{Fig:fig2}
\end{figure}

\begin{figure}
\includegraphics[width=0.9\linewidth]{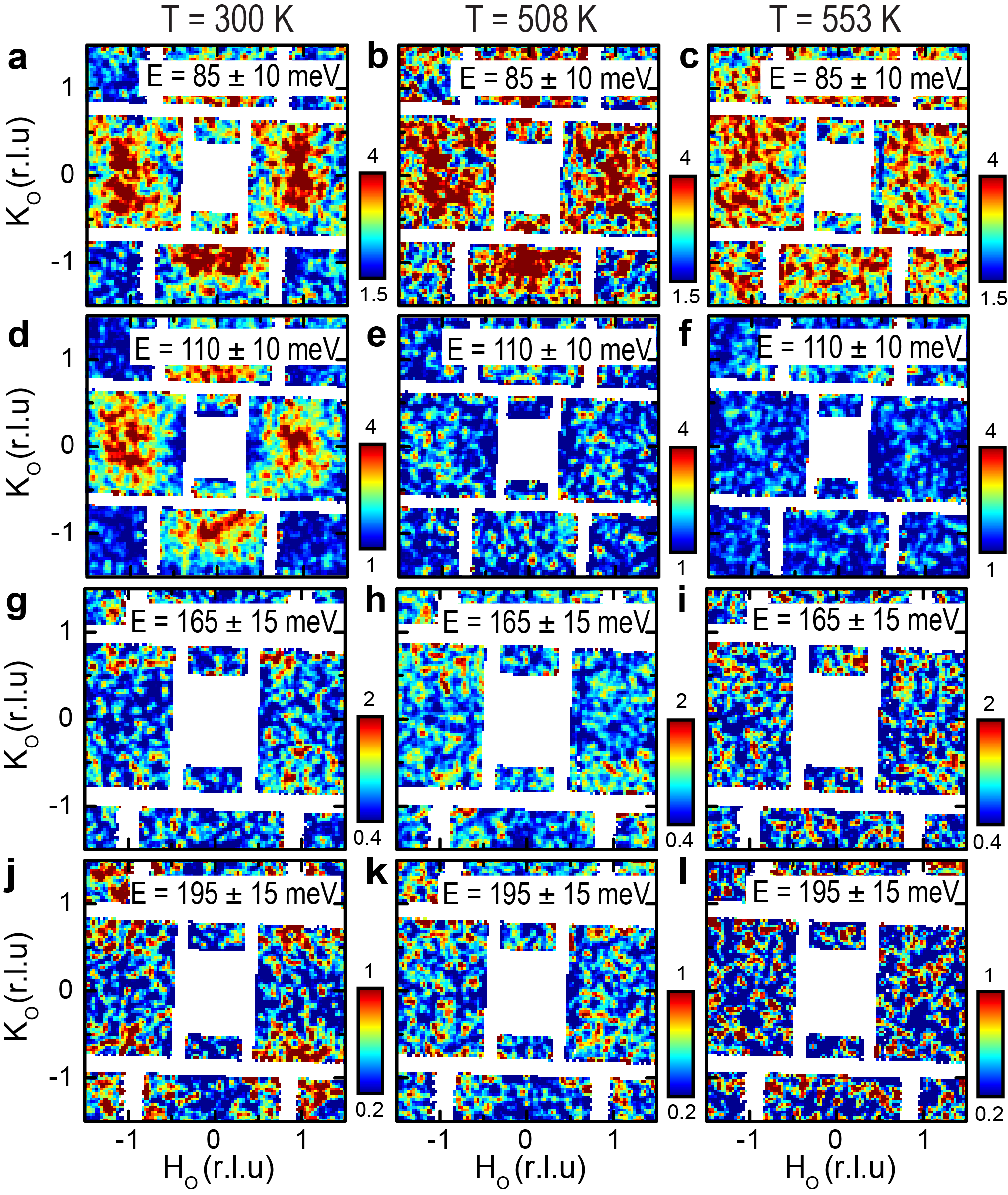}
\caption{(Color online) Wave vector and temperature dependence of optical 
spin waves and paramagnetic spin excitations at different energies for
Rb$_{0.8}$Fe$_{1.6}$Se$_2$. Spin excitations in the $[H_o,K_o]$ scattering plane at energies
(a,b,c) $E=85\pm10$, (d,e,f) $E=110\pm 10$, (g,h,i) $E=165\pm 15$, and (j,k,l) $E=195\pm 15$ meV.
The data in (a-f) and (g-l) are obtained with incident neutron beam energies $E_i=250$ and
440 meV, respectively, along the $c$-axis.
The left, middle, and right columns are identical spectra at 300 K, 508 K, and 553 K, respectively.
}
\label{Fig:fig3}
\end{figure}

Our experiments were carried out at the MAPS time-of-flight inelastic neutron scattering spectrometer at ISIS, Rutherford-Appleton Laboratory, UK as described previously \cite{harriger11}.  We grew single crystals of Rb$_{0.8}$Fe$_{1.6}$Se$_2$ using flux method \cite{mywang11}.
The chemical composition of these samples was determined from inductively coupled plasma analysis and found to be slightly different from those of previous
work \cite{mywang11}.
Below $T_N\approx 500$ K,
Rb$_{0.8}$Fe$_{1.6}$Se$_2$ forms an  Fe$_4$ block AF checkerboard structure with a $\sqrt{5}\times\sqrt{5}$ superlattice unit cell as shown in shaded area of Fig. 1a.
We define the wave vector Q at $(q_x, q_y, q_z)$ as $(H_o;K_o;L_o)=(q_xa_o/2\pi;q_ya_o/2\pi;q_zc_o/2\pi)$ rlu,
where $a_o=5.65$ and $c_o=14.46$ \AA\ are the orthorhombic cell lattice parameters
(green shaded area), for easy comparison with spin waves in BaFe$_2$As$_2$ \cite{harriger11,harriger12}.
Considering both left and right chiralities from the AF order, there are eight
Bragg peaks at wave vectors $(H_o,K_o,L_o)=(\pm 0.2+m,\pm 0.6+n,L_o)$ and
$(H_o,K_o,L_o)=(\pm 0.6+m,\pm 0.2+n,L_o)$ from the block AF structure, where $m,n=\pm 2,\pm 4, \cdots$, and $L_o=\pm 1,\pm 3, \cdots$ (Fig. 1b).
We coaligned $\sim$5 grams of single crystals of Rb$_{0.8}$Fe$_{1.6}$Se$_2$ (with mosaic $<3^\circ$) and 
loaded them inside a high temperature
furnace.
The temperature dependent AF Bragg peak and superlattice reflection associated with the $\sqrt{5}\times\sqrt{5}$
iron vacancy order disappear at $T_N=500$ K and $T_s=524$ K, respectively (Fig. 1c).
This indicates the vanishing magnetic and structure orders
 consistent with earlier results on other $A_y$Fe$_{1.6+x}$Se$_2$ \cite{weibao,fye,mwang11}.
Figure 1e shows the evolution of the acoustic spin waves with increasing temperature along the $[H_o,-0.5-0.5H_o]$ direction as shown
in the dashed line of Fig. 1b.  At 300 K, there are well-defined spin waves stemming from the block
AF ordered wave vectors (the upper panel, Fig. 1e).  Upon warming up to $T=1.02T_N=508$ K, paramagnetic spin excitations become much less well defined but still appear at the AF ordered wave vectors (the middle panel, Fig. 1e). Finally, on warming up to $T=1.06T_s=553$ K, paramagnetic spin excitations become featureless with no evidence for spin correlations at the AF ordering wave vectors (the bottom panel, Fig. 1e).

\begin{figure}
\begin{center}
\includegraphics[width=0.95\linewidth]{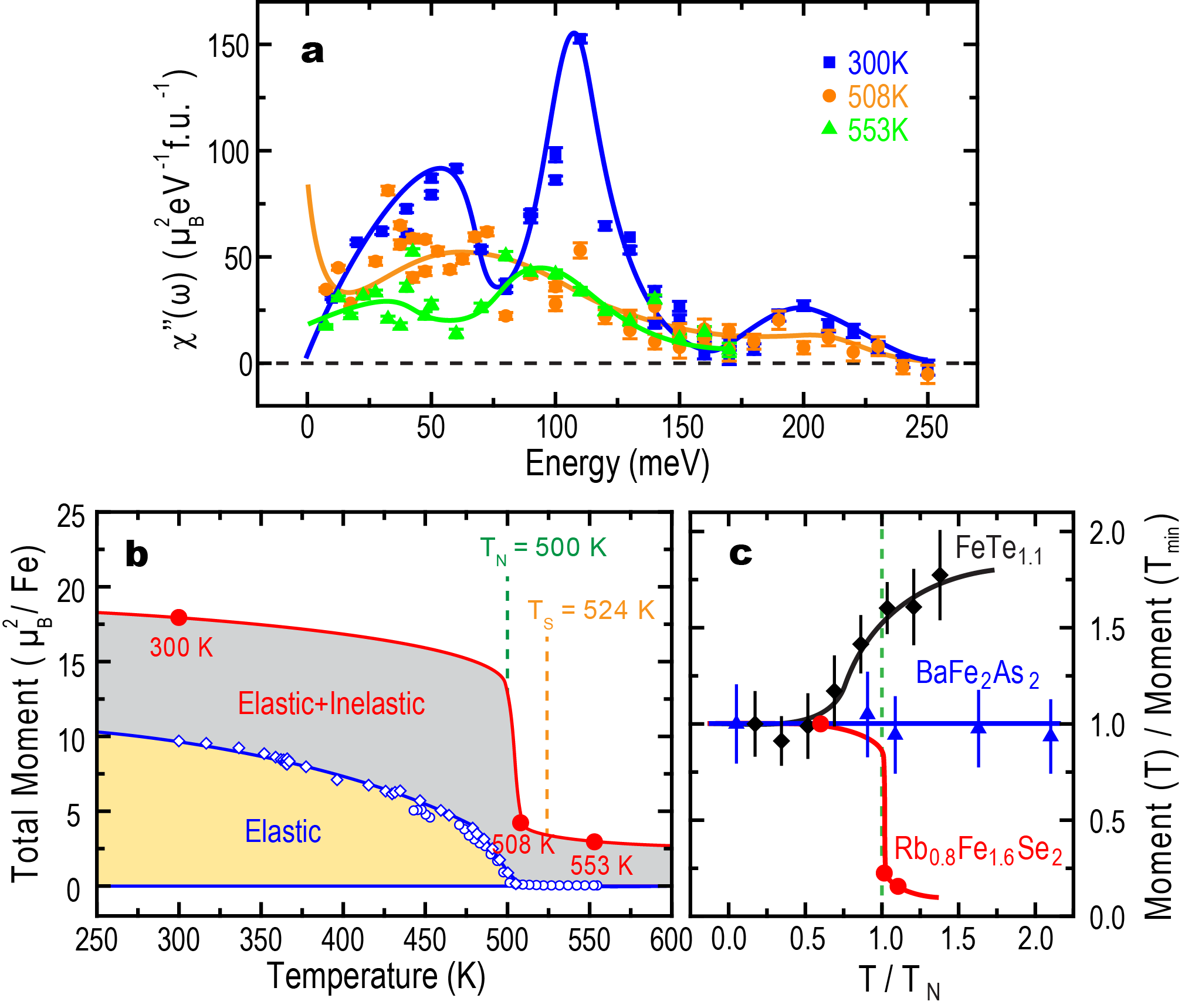}
\end{center}\caption{
(Color online)
(a) The energy dependence of the local susceptibility at 300 K, 508 K, and 553 K. The solid lines are guides to the eye.
(b) Temperature dependence of the energy integrated local susceptibility including both the static magnetic order
parameter and contribution from spin excitations, obtained by numerically summing up the data in (a). 
(c) Normalized total fluctuating moments $M(T)/M(T_{min}\ {\rm K})$ versus 
$T/T_N$ for Fe$_{1.1}$Te \cite{igor},
BaFe$_2$As$_2$ \cite{harriger12}, and Rb$_{0.8}$Fe$_{1.6}$Se$_2$.  The errors bars for Rb$_{0.8}$Fe$_{1.6}$Se$_2$ are smaller than
the size of the symbol.
}
\label{Fig:fig4}
\end{figure}

Figure 2 summarizes wave vector and temperature dependence of the low-energy acoustic
spin excitations in the $[H_o,K_o]$ plane from 300 K to 553 K.
At $T=0.6T_N=300$ K, spin waves are similar to the earlier results at 10 K \cite{mywang11}, 
having a spin anisotropy
gap at $E=6\pm 1$ meV and dispersing outward with increasing energy (Figs. 2a, 2d, 2g, 2j, 2m). 
In the AF ordered state, spin waves stem from the $\sqrt{5}\times\sqrt{5}$ in-plane wave vectors and $c$-axis wave vectors of $L=1,3,5$ \cite{mywang11}. 
On warming to $T=1.02T_N$, paramagnetic spin excitations become quasi two-dimensional with no $c$-axis modulations.
The spin anisotropy gap disappears and paramagnetic spin excitations move away from the $\sqrt{5}\times\sqrt{5}$ AF ordering positions
 for energies above $E=30$ meV (Figs. 2b, 2e, 2h, 2k, and 2n).
Upon further warming to above $T_s$ at $T=1.06T_s$, paramagnetic spin excitations become very broad in momentum space and move to
 the AF wave vector of BaFe$_2$As$_2$ instead of the block AF structure (Figs. 2c, 2f, 2i, 2l, and 2o).

Figure 3 shows the temperature dependence of the optical spin excitations.  For the low-energy optical spin excitations at $E=85\pm 10$ meV,
warming from 300 K (Fig. 3a) to 508 K (Fig. 3b) and 553 K (Fig. 3c) reduces the magnetic scattering intensity.  This can be seen from the broadening
of spin waves centered near $(\pm1,0)/(0,\pm1)$ positions at $300$ K to paramagnetic scattering 
essentially all wave vectors at $553$ K.
At $E=110\pm10$ meV, well-defined spin waves at 300 K (Fig. 3d) completely disappear at 508 K (Fig. 3e) and 553 K (Fig. 3f). 
At $165\pm15$ meV, there is no observable magnetic scattering at 300 K (Fig. 3g), 508 K (Fig. 3h), and
553 K (Fig. 3i).  Finally, spin waves centered near $(\pm 1,\pm 1)$ positions at $E=195\pm 15$ meV also vanish 
on warming from 300 K (Fig. 3j)
to 508 K (Fig. 3k) and 553 K (Fig. 3i).

Based on data in Figures 2 and 3, we construct in Figs. 1b, 1d, and 1f the evolution of spin waves to paramagnetic spin excitations in insulating
Rb$_{0.8}$Fe$_{1.6}$Se$_2$.  Comparing the result with dispersions of paramagnetic excitations in BaFe$_2$As$_2$ \cite{harriger11,harriger12}, where
high-energy spin excitations near the zone boundary are weakly temperature dependent for temperatures up to $2.1T_N$, we see that 
paramagnetic scattering in Rb$_{0.8}$Fe$_{1.6}$Se$_2$ behave much like a conventional local moment Heisenberg antiferromagnet, forming Lorentzian-like 
quasielastic peaks centered at $E=0$ \cite{tucciarone}.
To quantitatively determine the integrated magnetic moments and compare the outcome with those in Fe$_{1.1}$Te \cite{igor}
and BaFe$_2$As$_2$ \cite{harriger12}, we plot in Fig. 4 temperature dependence of the local dynamic
susceptibility for Rb$_{0.8}$Fe$_{1.6}$Se$_2$ \cite{msliu}.  For a local moment system with 
spin $S$, the total moment sum rule requires
$M_0=(g\mu_B)^2S(S+1)$ when magnetic scattering is integrated over 
all energies and wave vectors \cite{lorenzana}.  For iron in the $3d^6$ electronic state,
the maximum possible moment is $gS=4\ \mu_B$/Fe assuming $g=2$, thus giving $M_0=24\ \mu_B^2$/Fe.
In previous work \cite{mywang11}, we estimated that
the total moment sum rule is exhausted for Rb$_{0.89}$Fe$_{1.58}$Se$_2$ below $\sim$250 meV.
The energy dependence of the local susceptibility becomes progressively weaker on warming from 300 K to
508 K and 553 K (Fig. 4a).
Figure 4b shows temperature dependence of the ordered moment (open diamonds) \cite{weibao,fye,mwang11}
and integrated local susceptibility at three temperatures investigated (solid circles).
Consistent with earlier results \cite{mywang11}, we find that the total moment sum rule is almost exhausted
for Rb$_{0.8}$Fe$_{1.6}$Se$_2$ at 300 K, corresponding to a full moment of $gS=4\ \mu_B$/Fe with
$S=2$.  On warming to 508 K and 553 K, the total integrated moment drops dramatically, reflecting the fact
that our unpolarized neutron scattering experiment can only probe correlated magnetic excitations and
are not sensitive to wave vector independent paramagnetic scattering.
For comparison,
we note that the integrated magnetic spectral weight of Fe$_{1.1}$Te was found to
increase from the AF state to the paramagnetic state \cite{igor}, while the total integrated moment of BaFe$_2$As$_2$ remains essentially unchanged from
$T=0.05T_N$ to $T=2.1T_N$ \cite{harriger12}. To illustrate this point, we plot in Figure 4c 
the normalized total fluctuating moment ($M(T)/M(T_{min}\ {\rm K})$, where $M(T_{min}\ {\rm K})$ is integrated local moment in the lowest temperature
of the AF ordered state) as a function of $T/T_N$ for Fe$_{1.1}$Te \cite{igor},
BaFe$_2$As$_2$ \cite{harriger12}, and Rb$_{0.8}$Fe$_{1.6}$Se$_2$.  It is clear that 
temperature dependence of the fluctuating moment in Rb$_{0.8}$Fe$_{1.6}$Se$_2$ behaves 
differently from the other iron-based materials.

Comparing with iron pnictide BaFe$_2$As$_2$ and iron chalcogenide Fe$_{1.1}$Te, insulating
Rb$_{0.8}$Fe$_{1.6}$Se$_2$ appears to be a classic local moment Heisenberg antiferromagnet.
The lack of correlated high-energy paramagnetic spin excitations in Rb$_{0.8}$Fe$_{1.6}$Se$_2$ suggests
that electron correlation effects are smaller in Rb$_{0.8}$Fe$_{1.6}$Se$_2$,
contrasting to iron pnictides \cite{harriger12} and iron chalcogenide \cite{igor}.
This is also different from prototypical Mott insulators such as parent compounds of copper oxide
superconductors, where paramagnetic spin excitations above 100 meV are not expected to be different 
from spin waves below $T_N$ \cite{coldea}.  Our data thus suggests that insulating $A_y$Fe$_{1.6+x}$Se$_2$ is not a 
copper-oxide-like Mott insulator.  Alternatively, if insulating $A_y$Fe$_{1.6+x}$Se$_2$ is a semiconductor with an energy
gap of $\sim$500 meV opened below the $\sqrt{5}\times\sqrt{5}$ AF but not below the iron vacancy ordering temperature \cite{xwyan}, 
one would expect spin excitations to change dramatically from below to above $T_N$ but not significantly 
across $T_s$.  Although paramagnetic spin excitations in 
 the iron vacancy ordered state ($T=508$ K)
do appear at the $\sqrt{5}\times\sqrt{5}$ AF wave vectors for $E<20$ meV (Figs. 2b, 2e), higher energy acoustic and optical  
spin excitations are heavily damped and are 
sensitive to the magnetic but not to the iron vacancy order (Figs. 2 and 3).
This is consistent with the idea that insulating
Rb$_{0.8}$Fe$_{1.6}$Se$_2$ is an AF semiconductor \cite{xwyan}.
Finally, if magnetism in Rb$_{0.8}$Fe$_{1.6}$Se$_2$ arises from a combination of itinerant electrons and local moments due to
Hund's rule coupling similar to other iron-based materials \cite{dai,yzyou,wgyin,zpyin}, its paramagnetic spin excitations  should 
behave similarly as well.  Since paramagnetic spin excitations in iron chalcogenide and pnictides \cite{igor,harriger12}
are clearly different from those of Rb$_{0.8}$Fe$_{1.6}$Se$_2$  (Fig. 4c), our data adds to the 
debate on why superconductivity in $A_y$Fe$_{1.6+x}$Se$_2$ always appears near the $\sqrt{5}\times\sqrt{5}$ AF insulating phase \cite{weibao,fye,mwang11}, and which material is the true parent compound of $A_y$Fe$_{1.6+x}$Se$_2$ superconductors \cite{weili,zhao12}.

We thank Tao Xiang, J. P. Hu, and Guangming Zhang for helpful discussions.
The work at UTK is supported by the US DOE BES No. DE-FG02-05ER46202.
Work at the IOP, CAS is supported by the MOST of China 973
programs (2012CB821400, 2011CBA00110) and NSFC-51002180.

\end{document}